\documentclass[oldversion]{aa}

\usepackage{graphicx}
\usepackage{times}
\usepackage{dcolumn}
\usepackage{natbib}

\bibpunct{(}{)}{;}{a}{}{,}
\newcolumntype{.}{D{.}{.}{-1}}

\newcommand{\ks}{\mbox{$K_\mathrm{s}$}}

\begin{document}

\title{Host galaxies of bright high redshift quasars:\\ Luminosities
        and colours\thanks{Based on observations with the European Southern Observatory Very Large Telescope (ESO
        -VLT), observing run ID P69.B--0118(A).}}

\author{M. Schramm\inst{1}
        \and
        L.Wisotzki\inst{1}
        \and
        K. Jahnke\inst{2}
        }
\authorrunning{M. Schramm et al.}
\titlerunning{Host galaxies of high redshift quasars}

\institute{%
           Astrophysikalisches Institut Potsdam, An der Sternwarte 16, D-14482
           Potsdam, Germany
          \and
           Max-Planck-Institute for Astronomy, K\"{o}nigstuhl 17,
           D-69117 Heidelberg
           }
\offprints{M. Schramm, \email{mschramm@aip.de}}

\date{Received 19/02/2007 / Accepted 06/09/2007}

\abstract{We present the results of a near-infrared imaging study of high
         redshift ($z\sim 3$) quasars using the ESO-VLT. Our targets were
         selected to have luminosities among the highest known (absolute
         magnitude $M_B\la -28$). We searched for resolved structures
         underlying the bright point-source nuclei by comparing the QSO images
         with stars located in the same fields.  Two QSOs (HE~2348$-$1444 at
         $z=2.904$ and HE~2355$-$5457 at $z=2.933$) are clearly resolved in
         \ks, and with somewhat lower significance also in $H$; one object is
         resolved only in \ks. At these redshifts, $H$ and \ks\ correspond
         almost exactlly to rest-frame $B$ and $V$, respectively, with
         virtually no $K$-correction. We also report briefly the non-detection
         of some additional QSOs.  The detected host galaxies are extremely
         luminous with $M_V \sim -25$.  Their rest-frame $B-V$ colours,
         however, are close to zero in the Vega system, indicating substantial
         contributions from young stars and a stellar mass-to-light ratio
         below 1 (in solar units). Tentatively converting $M_V$ and $B-V$ into
         rough estimates of stellar masses, we obtain values of $M_\star$ in
         the range of several $10^{11}\:M_\odot$, placing them within the
         high-mass range of recent high-redshift galaxy surveys.   We present
         optical spectra and use \ion{C}{iv} line width measurements to
         predict virial black hole masses, obtaining typical values of
         $ {\bf M_\mathrm{bh}\sim 5\times10^{9}\,M_{\odot} }$. With respect to the
         known correlation between host galaxy luminosity
         $L_{V,\mathrm{host}}$ and $M_\mathrm{bh}$, our measurements reach to
         higher luminosities and redshifts than previous studies, but are
         completely consistent with them.  Comparing our objects with the
         local ($z\simeq 0$) $M_\mathrm{bh}$--$M_\mathrm{bulge}$ relation and
         taking also the low stellar mass-to-light ratios into account, we
         find tentative evidence for an excess in the 
         $M_\mathrm{bh}/M_\mathrm{bulge}$ mass ratio at $z\sim 3$.
          }

\maketitle

\section{Introduction}

Our understanding of quasar host galaxies has made rapid and significant
progress over the past decade. It is now clear that luminous quasars ($M_B \la
-24$) at low redshifts are mostly hosted by massive elliptical galaxies
\citep{McLure(1999),Dunlop(2003)}, irrespective of radio power.  Nuclear and
host bulge luminosities are correlated \citep{McLeod(1995)}, which is
naturally explained by the tight relation between the masses of supermassive
black holes and bulges \citep[e.g.,][and references therein]{Haering(2004)},
in particular if quasar black holes typically accrete and radiate at a fixed
fraction of the Eddington rate \citep{McLeod(1999)}.

High redshift quasar hosts have also received increased attention lately, as
they provide one avenue to study the assembly and evolution of massive
galaxies, in particular in relation to the growth of their central black
holes. Yet, quantitative knowledge of QSO host properties at redshifts well
beyond $z\ga 1$ is still very limited, mainly due to the fact that optical
imaging only gives access to the rest-frame ultraviolet spectral range where
the QSO nuclei outshine any host galaxy.  The improvements in near-infrared
instrumentation have opened also this window, first with the \emph{Hubble
Space Telescope} \citep[][]{Kukula(2001),Ridgway(2001)}, followed by some
ground-based experiments
\citep{Falomo(2004),Falomo(2005),Kuhlbrodt(2005)}. Very useful insights have
also been obtained from HST imaging of gravitationally lensed QSOs
\citep{Peng(2006)3}.

These studies were all limited to single band observations, typically sampling
the rest-frame optical wavelength range. Largely focussed on solving the
challenging problem of resolving the QSO images and deblending the nuclear and
underlying extended galaxy components, their main result is that QSOs are
hosted by fairly luminous galaxies already at high $z$. More quantitatively,
\citet{Kukula(2001)} found essentially no evolution between $z\simeq 2$ and
$z\simeq 0$ in typical host luminosities, for radio-quiet QSOs at given
absolute nuclear magnitude of $M_B \sim -24$. This trend has been confirmed by
more recent studies based on larger samples
\citep[e.g.][]{Peng(2006),Peng(2006)3}.

Unfortunately, additional information such as morphological properties or
colours are even harder to come by than single band luminosities. Especially
colours, or more generally speaking, spectral data are very scarce already at
much lower redshifts, and significant improvements have only been obtained in
the last few years, at $z\la 0.2$ \citep{Jahnke(2004)} and at $z\simeq 0.7$
\citep{Sanchez(2004)}.  These studies have demonstrated that there is strong
evidence for the presence of of blue, young stars in low-redshift
QSO host galaxies, including those that appear to be quiescent ellipticals.

Clearly, one would expect recent or ongoing star formation to be even more
prevalent in galaxies at high redshifts. Indeed we found substantial UV
emission from the hosts of moderate luminosity QSOs at $z\simeq 2.2$ using
HST/ACS imaging \citep{Jahnke(2004)2}. But to our knowledge, no measurements
of rest-frame optical colours have yet been published for high-redshift QSO
host galaxies.

In the present paper we report on the results of a pilot study, aiming at
measuring host galaxy luminosities in two different near-infrared bands using
the ESO-VLT. Our targets are radio-quiet, very high luminosity QSOs at $z\sim
2$ and $z\sim 3$. We briefly describe our strategy to decompose the QSO
images into their nuclear and host contributions, and we discuss the results
in the context of recently obtained scaling relations. Throughout this paper 
we assume a flat cosmology with $H_0=70\mathrm{ km\,s}^{-1}\,\mathrm{Mpc}^{-1}$, $\Omega_M=0.3$ and
$\Omega_{\Lambda}=0.7$. All magnitudes are expressed in the Vega system.

\section{Observations and data reduction}

\subsection{Target selection}

\begin{figure}
\begin{center}
  \includegraphics[clip,angle=-90,scale=0.47]{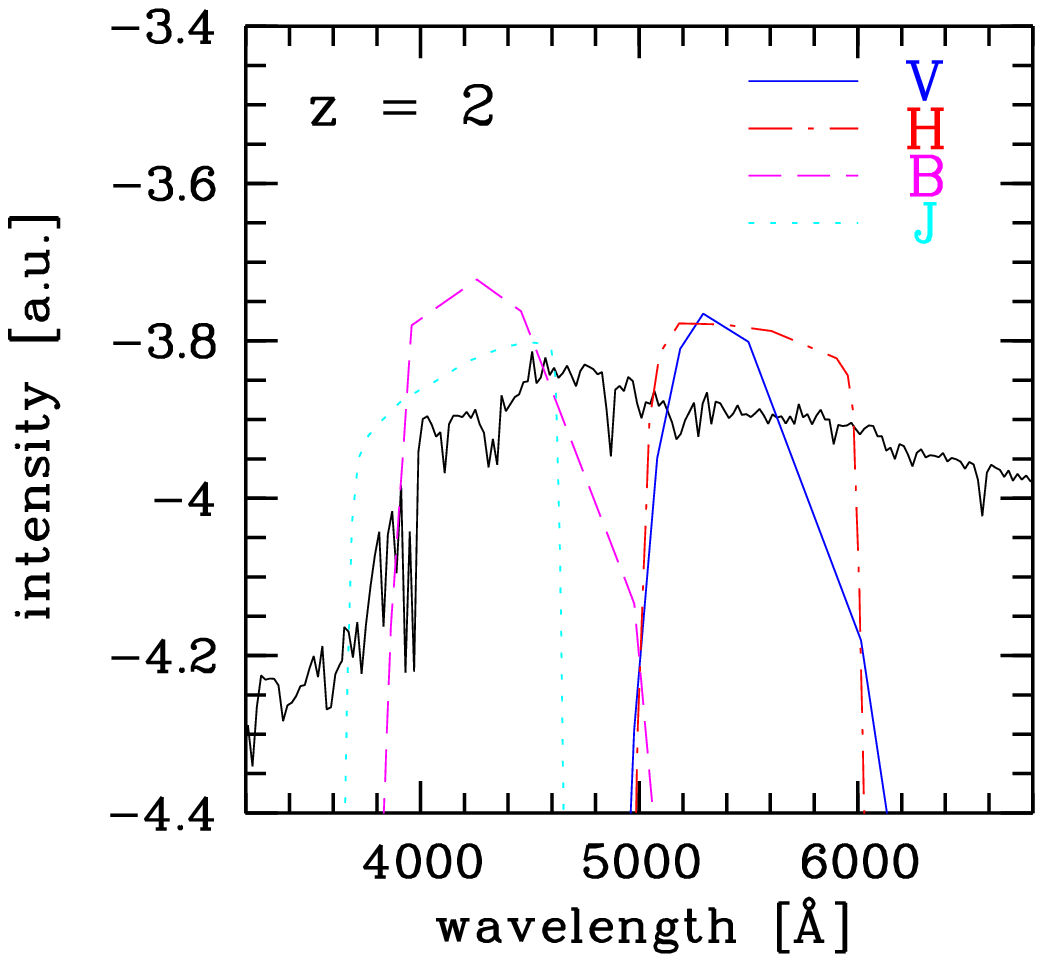}
  \includegraphics[clip,angle=-90,scale=0.47]{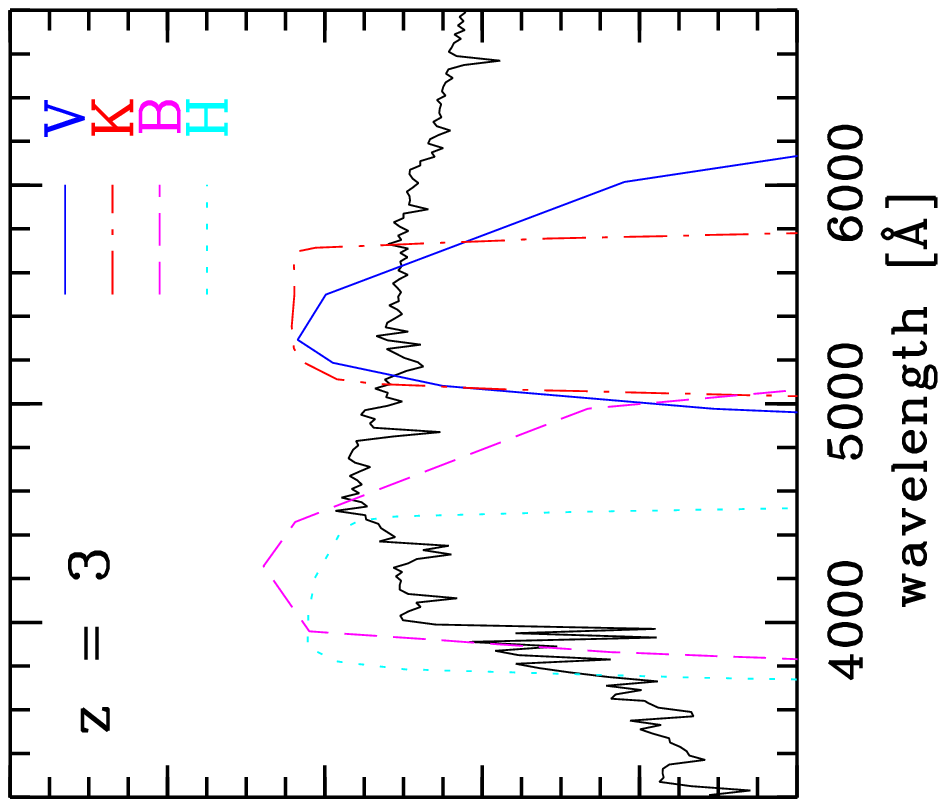}
 \caption{Position of the corresponding filter profiles at $z=2$ and $z=3$.
          In both panels, the spectral template represents a single stellar
          population model of a 1 Gyr old galaxy.}
\label{filter}
\end{center}
\end{figure}

Our aim was to observe quasars with the highest possible luminosities, which
we expect to reside in the most massive bulge-dominated galaxies.  Using the
Hamburg/ESO Survey for bright QSOs \cite[HES;][Wisotzki et al., in
prep.]{Wisotzki(2000)} as input catalogue matched this aim.  Our QSOs were
selected to have redshifts $2.6<z<3.0$ so that the rest-frame $B$ and $V$
bands were approximately reproduced by the standard near-infrared filters $H$
and \ks, respectively. We also added a few QSOs with $1.8 < z < 2$, where $J$
and $H$ correspond to rest-frame $B$ and $V$, respectively.  Figure
\ref{filter} illustrates the good match between rest-frame and observed-frame
filter bands. We verified by visual inspection in the Digitized Sky Survey
that there was at least one star of comparable brightness within $\sim 1$
arcmin from each QSO, which later would be usable for calibrating the
point-spread function. Table~\ref{T:obs} lists the small sample of objects
for which we obtained useful data.

\subsection{Observations}

The sample was observed with the ESO-VLT in service mode (thus spreading over
several nights) between May and July 2002.  The instrument used was the
Infrared Spectrograph and Array Camera (ISAAC), mounted on the Nasmyth B focus
of UT1 (Antu) of the VLT.  Because of technical problems with the short
waveband (`Hawaii') arm, only the long waveband arm was available. This was
equipped with a SBRC Aladdin 1024$\times$1024 pixel InSb array providing a
pixel scale of $0\farcs 148$ pixel$^{-1}$ for a total field of view of
$151''\times 151''$. The data were obtained in observing blocks consisting of
sequences of 12 exposures in $J$, 15 in $H$ and 30 in the \ks\ band. In order
to account for the rapidly varying NIR night sky, the telescope was moved
between successive exposures following a multi-point dither pattern along a
sequence of predefined quasi-random grid points. Exposure times per dither
pointing were 150, 120 and 60 s in $J$, $H$ and \ks, respectively, each split
into several short integrations. The final net
integration times per target and filter band are listed in Table \ref{T:obs},
along with the effective seeing measured in the coadded images.  All
observations were obtained under photometric conditions.

\begin{table*}
      \caption[]{Target List and Log of Observations.}
         \label{T:obs}
\begin{center}
\begin{tabular}{ccccccccccc}
\hline\noalign{\smallskip}
    source       & RA          & Dec      & $z$   & $B$    &
    \multicolumn{3}{c}{exp. time [min]} & \multicolumn{3}{c}{seeing [arcsec]} \\
                 &             &          &       &        &                   $J$  &   $H$ & \ks                    & $J$       &   $H$    &    \ks \\
\noalign{\smallskip}\hline\noalign{\smallskip}
    HE 2329$-$4115 & 23 31 50.5  & $-$40 58 53 & 1.850 & 18.7  & 60 & 30 &--& 0.75 & 0.70 & --  \\
    HE 2213$-$3722 & 22 16 30.8  & $-$37 07 32 & 1.920 & 17.6  & 30 & 60 &--& 0.71 & 0.73 & --  \\
    HE 2149$-$0436 & 21 51 50.1  & $-$04 22 19 & 2.643 & 17.3  & -- & 60 &60& -- & 0.58 & 0.71  \\
    HE 2355$-$5457 & 23 58 33.5  & $-$54 40 42 & 2.904 & 17.8  & -- & 90 &90& -- & 0.73 & 0.70  \\
    HE 2348$-$1444 & 23 51 29.8  & $-$14 27 57 & 2.933 & 17.3  & -- & 60 &60& -- & 0.65 & 0.65  \\
\noalign{\smallskip}\hline
\end{tabular}
\end{center}
\end{table*}

\subsection{Data reduction}

The images were reduced using ESO-MIDAS and the \texttt{eclipse} software
package also provided by ESO, in particular the eclipse-based ISAAC reduction
pipeline recipe \texttt{lwjitter}. The basic steps involved (i) flat fielding
using twilight flats, (ii) estimate of a sky frame by median filtering of the
individual frames in the stack, after rescaling to account for temporal sky
level variations, (iii) sky subtraction, (iv) determination of the angular
offsets between frames, using an auto-correlation technique, (v) final image
registration and coaddition for each stack of images.

As all observations in the NIR, our data are totally dominated by the sky
background. It turned out that the background removal by the reduction
pipeline was not quite good enough for a detailed analysis of very low surface
brightness features. We therefore determined a local background correction for
each object separately by creating a curve of growth.  This generally worked
satisfactorily, although the data sometimes suffered from spatial background
features varying with time, which we could not correct for. The main effect of
such variations was an increase of the statistical uncertainty of the local
background correction. We quantified this uncertainty by extracting several
small control images of blank sky in arbitrary positions, which were then
treated in the same way.

\subsection{Photometry and absolute magnitudes}

Flux calibration was based on observations of standard stars from
\citet{Persson(1998)}. Atmospheric extinction coefficients were taken from the
ESO web pages. The derived zero points from various stars turned out to be
highly consistent, with a median internal error of 0.003 mag.

Conversion of apparent into absolute magnitudes requires the knowledge of a
$K$-correction (including the band pass stretching). By design of our
experiment, the spectral corrections are very small (cf.\ Fig. \ref{filter}).
We followed in principle the approach described by \citet{Jahnke(2004)}.  For
the K-correction of the nucleus we used an average quasar SED, for the host
galaxies we calculated K-corrections from single stellar population
models with stellar ages ranging between 0.1--1 Gyr.  The resulting absolute
magnitudes, however, are almost independent of the assumed spectral energy
distributions.

\section{Analysis}

\subsection{Point spread function}

The detection of faint extended fuzz underlying a bright point source is
notoriously difficult and requires foremost an accurate knowledge of the
effective point spread function (PSF) of the optical system. All quasars in
the sample were selected to have at least one suitable PSF star in their
vicinities. Unfortunately, a detailed analysis of stellar profiles within the
full ISAAC field of view revealed significant spatial variations of the PSF,
reminiscent of optical aberrations. Even worse, the pattern clearly varied
also in time, possibly due to focus variations, and not two images shared the
same pattern.  While in principle it is possible to model such distortions
using several stars in the field of view \citep[e.g.,][]{Kuhlbrodt(2004)}, the
number of stars at these high Galactic latitudes was mostly too small. 
Some of our images thus became essentially useless, with the PSF stars displaying
a broader radial profile than the QSOs. Such data had to be discarded
entirely, and they are not discussed any further in this paper.

In the end, we used only exposures where the PSF appeared to be reasonably
stable within the central regions. In our best cases we had two stars close to
the quasar that were bright enough to be useful as PSF calibrators; here we
used either an average PSF for each given image, or each star individually
when explicitly quantifying the effects of PSF uncertainty.  In particular we
adopted the difference image of the two stars as a rough guess of the
additional `noise' contributed by the PSF uncertainty.

\subsection{PSF subtraction} \label{sec:psfsub}

PSF subtraction is a well-established technique for the analysis of QSO host
galaxy images. It has the big advantage that a minimum of assumptions about
the nature of the host is entering the analysis. Here we employed it with the main purpose of
performing a conservative test for the `null hypothesis', namely that
no resolved host galaxy is detectable underneath the QSO point source.

Once a PSF was established, it had to be shifted to the location of the QSO,
scaled, and subtracted. To minimize shifting residuals we used a subpixel grid
for the centroid position and found the best location from $\chi^2$
minimisation. As a scaling criterion we demanded that the summed residual flux
in a small aperture of 2 pixels radius around the centroid vanished after PSF
subtraction. Unresolved sources can thus be robustly identified, as any
residual flux after PSF subtraction should be consistent with PSF mismatch
errors. On the other hand, it is obviously a very conservative criterion for
resolved sources implying a significant oversubtraction in case a host galaxy
should be present. However, a significantly detected extended source should
still yield systematically positive residuals at larger radii. Only those
detections were taken as significant that show positive residuals for most of
the bright stars that are located in the vicinity around the QSO and contain
at least 5~\% of the QSO flux after oversubtraction correction. In such cases
it is even possible to approximately account for the oversubtraction by
applying a correction factor derived from simulations (see
\citealt{Jahnke(2004)2} for a more extensive discussion of this procedure).
Briefly, we performed a series of such simulations where artificial host
galaxies were added to one of the PSF stars. After adding appropriate noise we
took a second star in each field as PSF calibrator which was shifted, scaled,
and subtracted.  This provided us with a rough estimate of the systematic PSF
oversubtraction. The model host galaxies in this exercise were assumed to be
smooth and circularly symmetric, with a surface brightness distribution
described by a de Vaucouleurs profile of effective radius of 3, 5 and 10~kpc,
respectively. This is slightly more compact than present-day high-mass
elliptical galaxies, but plausible given the high redshifts involved and given
earlier results \citep[see also ][]{Kuhlbrodt(2005)}. While typical values of
the residual flux are of the order of a few percent we determined 
oversubtraction correction factors of 0.5--1 mag for the \ks\ band and
somewhat higher values of 1--1.5 mag in case of the $H$ band.
We obtained
similar correction factors if an exponential disk is assumed instead of a de
Vaucouleurs profile as galaxy model. We find the apparent host galaxy
magnitudes to be fainter by 0.2--0.5 mag and stronger residuals at the same
time. After the oversubtraction correction was applied, all nuclear and host
galaxy magnitudes agree within 0.2 mag with the values determined by our
two-dimensional modelling technique. A similar procedure as described
above was applied for the non-detections to determine upper limits of
the host luminosity. Here we created series of quasars images by adding a
synthetic galaxy component of 3, 5 and 10~kpc, respectively, to the given PSF.
We varied the nuclear/host flux ratio until we clearly resolved the galaxy in
the radial surface brighness profile. We adopted as upper limits when the
quasar surface brightness deviated from the PSF profile by 3$\sigma$ based on
shot noise error bars determined in a ring excluding the central pixels.

\subsection{Two-dimensional modelling} \label{sec:mod}

A more advanced deblending treatment is based on the explicit assumption that
the surface brightness profile of an observed host galaxy should roughly
follow a simple analytical prescription. It is then possible to fit a model of
the two-dimensional light distribution to the data, consisting of a
superposition of a scaled PSF plus an analytical galaxy model convolved with
the PSF. We used our own code PAMDAI \citep{Kuhlbrodt(2004)} which in the past
was successfully applied to low-redshift QSO hosts
\citep[e.g.,][]{Jahnke(2004)}. The code determines the best-fit
parameters by $\chi^2$ minimisation employing a modified downhill simplex
algorithm. 

We soon realised that the signal of any possible host galaxy was too weak in
our data to permit simultaneous constraining of the shape and the flux level
of the galaxy model. We therefore made the following additional assumptions:
The host galaxy was assumed to follow a de Vaucouleurs law with zero
ellipticity and with a preset effective radius $r_\mathrm{eff}$. When
assuming an exponential law instead, we find the host magnitudes to be fainter
by 0.3--0.4 mag.  Thus, only the fluxes of the nucleus and galaxy 
had to be determined by the fit. With these provisions the fits converged well
for those objects where the PSF subtraction indicated a resolved source.  It
turned out that the $\chi^2$ values resulting from assuming different fixed
values of $r_\mathrm{eff}$ came out very similar as long as realistic
$r_\mathrm{eff}$ values were chosen. It is thus not possible to significantly
constrain the scale length of the host galaxy from our data, which is perhaps
not too surprising as our angular resolution of $0\farcs 7$ corresponds to
$\sim 5$~kpc at $z=3$. However, the estimated host galaxy fluxes do not depend
sensitively on the correct scale length. This fact is known for long
\citep{Abraham(1992)}, and we confirm it to be valid for our present data,
also through extensive simulations.

We computed a two-dimensional model only in those cases where the
conservative PSF subtraction revealed positive residuals. We then adopted the
best-fit PSF scaling factor to obtain a best-guess PSF-subtracted image of the host galaxy.

\begin{figure}
\begin{center}
  \includegraphics[clip,angle=0,scale=0.6]{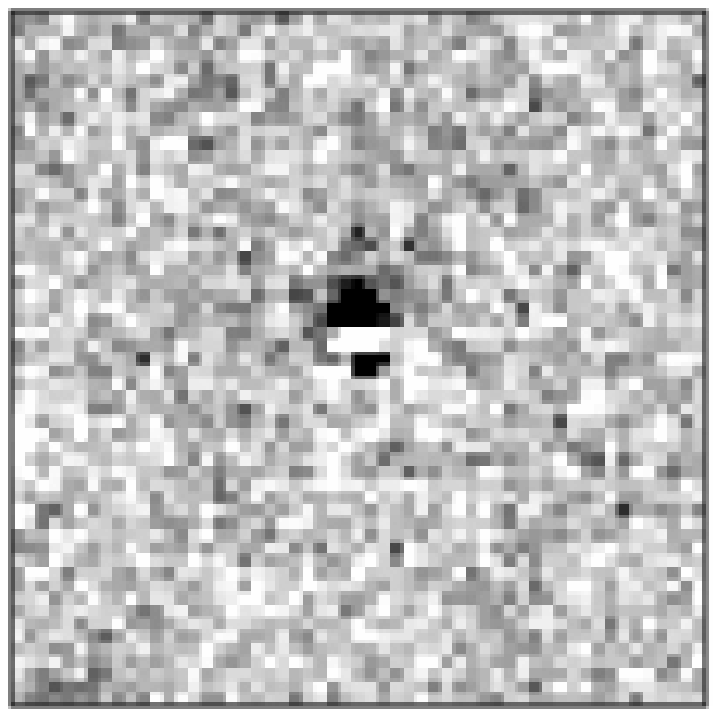}
  \includegraphics[clip,angle=0,scale=0.6]{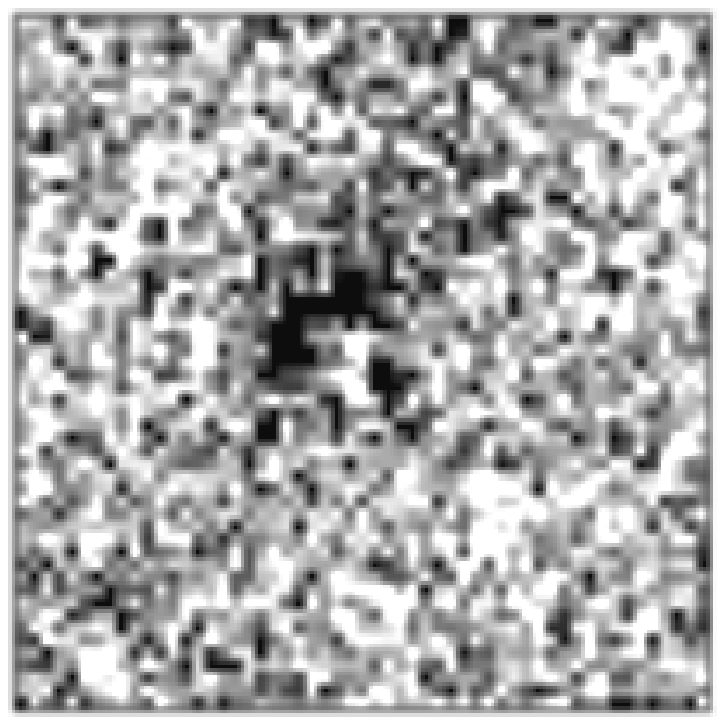}
 \caption{Residuals after subtracting one PSF star from another star for
  HE~2355$-$5457 (left) and HE~2348$-$1444 (right). We used the two stars marked as S1
  and S2 in Fig.~\ref{qso_im}. Cuts are chosen the same as for Fig.~\ref{qso_im}.}
\label{psf_psf}
\end{center}
\end{figure}

\section{Results}

\begin{figure*}
\begin{center}
\includegraphics[clip,angle=0,width=509pt]{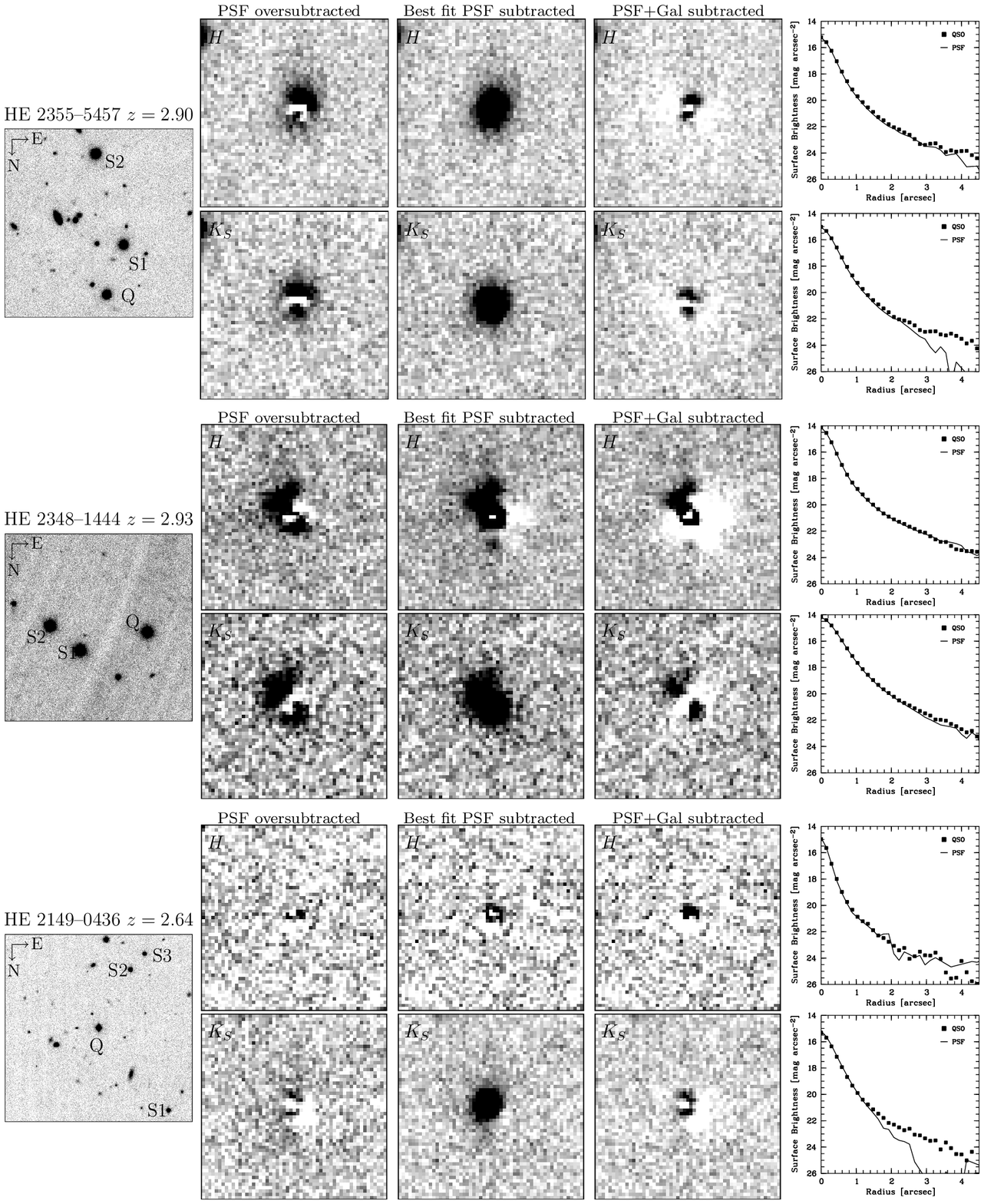}
  \caption{Host galaxy detection of HE~2355--5457 (upper panel), HE~2348--1444
    (middle panel) and HE~2149--0436 (bottom panel). The left-hand images have
    a scale of $1'\times 1'$ and show the QSO with PSF calibrator stars; for
    each QSO, $H$ band in the top row, \ks\ band below. All other images are
    $8''\times 8''$ postage stamps zoomed on the QSO. The three middle panels
    show the residual host galaxy light after PSF subtraction; middle left:
    based on conservative PSF subtraction; center: based on the best-fit PSF
    scaling; middle right: residuals left after subtracting both the best-fit
    PSF and an azimuthally symmetric model host galaxy. In the right hand
    panels we show the radial surface brightness profiles of the QSO and
    corresponding best-fit PSF.}
\label{qso_im}
\end{center}
\end{figure*}

We succeeded to resolve the host galaxies of three QSOs from the higher
redshift subsample, two of them in both $H$ and \ks\ bands, and one only in
\ks. We failed to detect the hosts of two QSOs from the lower redshift
subsample. We now briefly discuss the sources individually. A summary is
given in Table~\ref{T:highmag1}, images of the detected QSOs are shown
in Fig.~\ref{qso_im}.

\paragraph{HE 2355$-$5457: }
This field has two nearby stars of comparable brightness as the QSO located
within less than $45''$ distance from the QSO. There is no evidence for
significant PSF variations over the central regions of the field.  Subtracting
a composite scaled PSF from the QSO left a strong excess of positive pixel
values outside of the very central zone, in both $H$ and \ks\ bands. We tested
for the reality of this effect by subtracting one PSF star separately from
either QSO or the other star. The difference of QSO$-$PSF showed consistently
a $10\times$ higher residual flux than the difference of PSF1$-$PSF2. Visual
inspection of the PSF-subtracted QSO image showed that this excess in the
quasar profile was not caused by a close companion, or by a strongly
asymmetric feature, but that it was due to more or less azimuthally symmetric
low surface brightness emission centered on the QSO. We conclude that our
conservative null hypothesis of the QSO having no detectable host galaxy has
to be rejected, with high significance for the \ks\ band, but also (with
somewhat lower significance) for the $H$ band.  However, with the galaxy
contributing only $\sim 5$~\% of the total uncorrected \ks\ band flux, the
host is still much fainter than the central nucleus. In the $H$ band this
fraction is even slightly less than 3~\%. After applying a statistical
correction for oversubtraction (see Sect.\ \ref{sec:psfsub}) these values
increase to $\sim 10$~\% and $\sim 8$~\% for the \ks\ and H band,
respectively. We then computed two-dimensional model fits (with fixed scale
lengths as discussed above), and obtained integrated \ks\ and $H$ band fluxes
for the host galaxy that are in good agreement with the corrected estimates
based on PSF-subtraction. This makes HE 2355$-$5457 the QSO with the highest
significance detection in the whole sample. The results are documented in the
upper panels of Fig.\ \ref{qso_im}.

\paragraph{HE 2348$-$1444: }
The situation with this QSO is similar to the previous object: We have two
bright stars close to quasar which show very similar surface brightness
profiles. We again find positive PSF-subtraction residuals in both bands which
were confirmed by the same test procedure as described for HE
2355$-$5457. However, the fraction of an extended flux component is even lower,
only 1-2~\%, or 6~\% and 5~\% in \ks\ and
$H$, respectively, after correction for oversubtraction. Notice that 
HE 2348$-$1444 is the brightest quasar in our sample. In this object
the residual flux is strongly asymmetric (see Fig.\ \ref{qso_im}).
This feature becomes particularly evident after subtracting the
azimuthally symmetric two-dimensional model fit. At the same time,
the lacking symmetry makes the extended nature of the object unconspicuous in
the radial profiles (it is in fact not visible in the $H$ band profile). 
Nevertheless, we believe that since the feature appears as highly consistent
in both bands, its reality can be safely assumed. We conclude that also in
this QSO we have detected the host galaxy, and that there is evidence for
a highly disturbed morphological structure, most likely due to gravitational
interaction or merging.

\paragraph{HE 2149$-$0436: }
The field contains several (mostly faint) stars showing that PSF variations
for the whole field of view are rather large. For both bands we find positive
as well as negative residuals after subtraction of individual stars extracted within
~30'' around the QSO. In case of the \ks\ band, the number of sufficiently
bright field stars was acceptable to create an analytical model of the spatial
PSF variations (similar to the procedure by \citealt{Kuhlbrodt(2004)}). 
Using the best-fit analytical PSF prescription at the QSO position, we
detected an extended source after PSF subtraction.
The amount of uncorrected flux is 6~\% and increases to 13~\% after
oversubtraction correction. The image quality of the $H$ band data was
unfortunately inferior to the \ks\ data, and we did not succeed in resolving
the host galaxy. In the following we only quote a 3$\sigma$ upper limit of the 
host galaxy luminosity, propagating into corresponding lower limits 
of the rest-frame $B-V$ colour.

\paragraph{HE 2329$-$4115 and HE 2213$-$3722: }
While the $J$ band data of these $z\sim 2$ QSOs turned out to be
useless, we found both QSOs to be marginally resolved in the $H$ band,
but below our formal significance criterion. We therefore only quote
3$\sigma$ upper limits in Table~\ref{T:highmag1}.

\begin{table*}[tb]

\begin{center}
{\small
\caption[]{ Integrated apparend magnitudes and absolute magnitude estimation
  for the observed QSOs. The results are quoted for an
  fixed effective radius of 5 kpc.
}
\label{T:highmag1}

\begin{tabular}{lrrrrrrrrr}

\hline\noalign{\smallskip}
Object      &$H^\mathrm{host}$ & $H^\mathrm{nuc}$& $\ks^\mathrm{host}$ & $\ks^\mathrm{nuc}$ &  $M_B^{\rm{nuc}}$ & $M_B^{\rm{host}}$ & $M_V^{\rm{nuc}}$ &$M_V^{\rm{host}}$   \\
             &                   &                 &                      &                   &                   &                   &                  &                    \\
\noalign{\smallskip}\hline\noalign{\smallskip}
HE~2149$-$0436    &$>$18.0       &16.3      &18.0      &15.9  & $-$27.4 & $>-$25.6 & $-$27.4 &$-$25.4\\
HE~2355$-$5457    & 18.0 &15.8& 17.8 & 15.5& $-$28.2 & $-$25.8 & $-$28.2  &$-$26.0\\
HE~2348$-$1444    & 17.8 & 14.8 & 17.4 & 14.2 &$-$29.3 & $-$26.4 & $-$29.5 &$-$26.3\\
HE~2213$-$3722   & $>$18.5  & 16.6&     &        &&&$-$26.7&$>-$24.8&    \\
HE~2329$-$4115   &  $>$17.8 & 16.9&     &        &&& $-$26.3&$>-$25.4&      \\

\noalign{\smallskip}\hline
\end{tabular}

}
\end{center}
\end{table*}

\section{Discussion}

\subsection{Host galaxy luminosities}

The fluxes of the detected host galaxies were determined by integrating 
over the PSF-subtracted images, where the PSF scaling factor was given by
the best-fit analytic model. Since we did not allow the half-light radii to vary
during the fit, the resulting fluxes depend somewhat on the assumed size of
the galaxy.  We found that our adopted range for $r_{1/2}$ from 3~kpc to
10~kpc corresponds to a systematic uncertainty of $\pm 0.2$~mag, regardless of
the used filter. In those cases where only upper limits are quoted, the
conservatively PSF-subtracted images were used.

After applying distance modulus and $K$ corrections as discussed above,
we obtained absolute magnitudes in the
rest-frame $V$ and $B$ bands. We reiterate that our measurements essentially
sample the objects at these rest-frame wavelengths, and that $K$ corrections
are almost negligible. The resulting absolute magnitudes are listed in
Table~\ref{T:highmag1}. Notice that the quasars in our sample are so powerful
that despite the high nuclear-host ratio of $\ga 10$, the hosts are still
extremely luminous galaxies.

\begin{figure}
\begin{center}

 \includegraphics[clip,angle=-90,width=88mm]{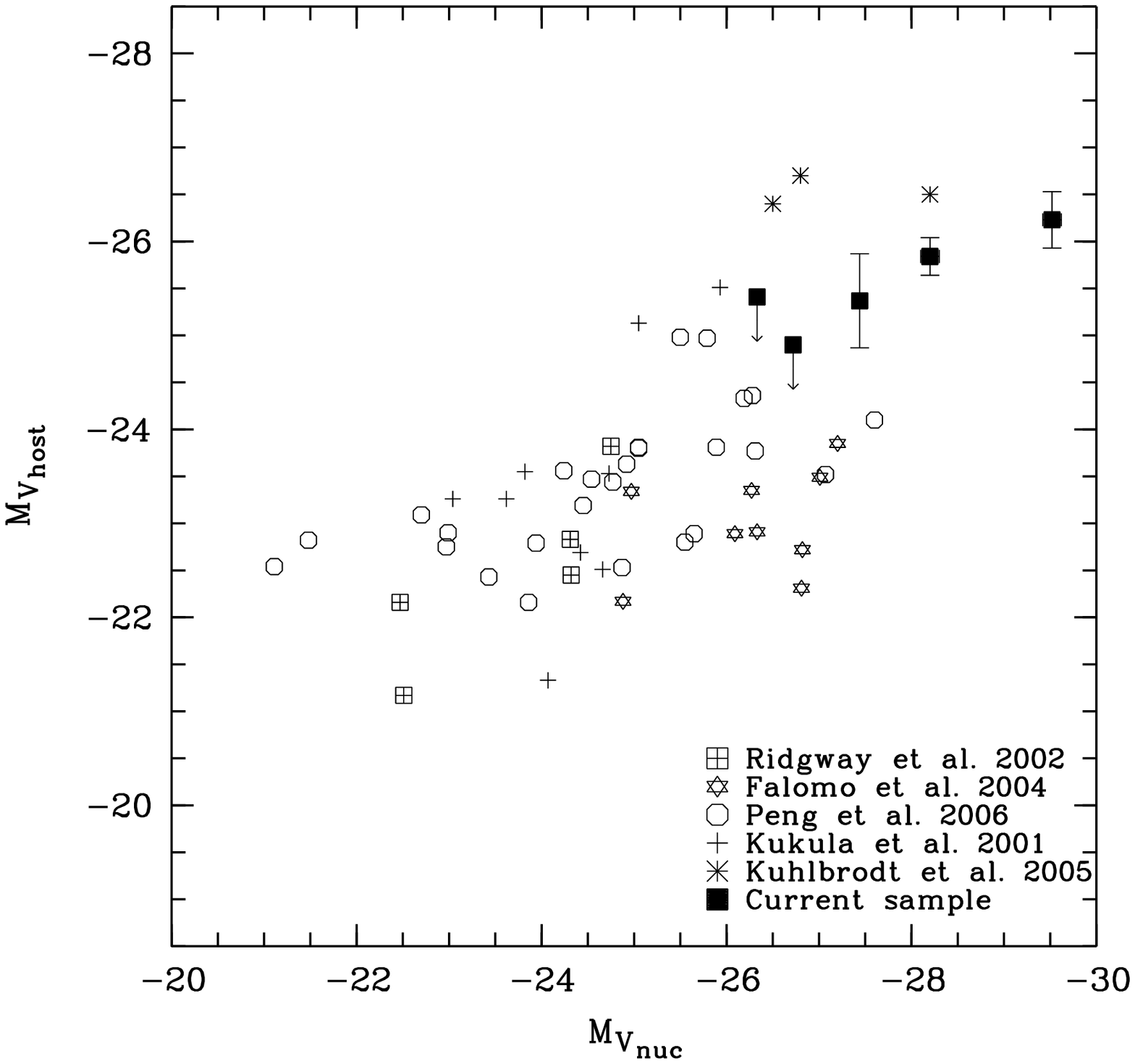}
 \caption{ Relation between nuclear and host galaxy luminosities for
high-redshift ($z > 1.4$) QSOs. Our data are represented by the filled
squares. For comparison we plott several results from other recent studies:
\citep{Ridgway(2001),Kukula(2001),Falomo(2004),Kuhlbrodt(2005),Peng(2006)},
all numbers converted to our set of cosmological parameters.  Objects
radiating at constant fraction of their Eddington luminosity would lie on a
diagonal line in this diagram.}
\label{mvh_mvn}
\end{center}
\end{figure}

Figure~\ref{mvh_mvn} compares our measurements to other recent studies of
high-redshift QSO host galaxies. We restrict this comparison to near-infrared
studies where $M_V$ of the host galaxy can be estimated without major spectral
extrapolation. While evidently there are very few detections for
high-luminosity QSOs, our data line up well with the published observations.
As a whole Fig.~\ref{mvh_mvn} suggest that QSO nuclear and host galaxy
luminosities are correlated also at high redshifts, very similar to the
finding by \citet{Floyd(2004)} at low $z$.

QSOs radiating at a fixed fraction of the Eddington limit are distributed 
along a diagonal in this diagram \citep{McLeod(1999)}, assuming that black
hole masses are proportional to host bulge luminosities 
\citep{Magorrian(1998)}. Under these premises, our objects appear to have
similar Eddington ratios as less luminous QSOs at redshifts $z\ga 2$ (and in
fact also similar to low-$z$ QSOs). This very qualitative reasoning can
be quantified if mass estimates are available, for both
black holes and stellar bulges. In the following we attempt to estimate
both these quantities for our $z\sim 3$ QSOs.

\subsection{Black hole masses}

Several recent investigations have demonstrated that one can estimate the
masses of black holes in active galactic nuclei from single-epoch spectra of
broad emission lines \citep{Vestergaard(2002),Vestergaard(2006),McLure(2002)}. This approach
assumes the broad-line region (BLR) to be in approximate virial equilibrium
\citep{Peterson(2000)} and adopts an empirical luminosity-BLR size scaling
relationship \citep{Kaspi(2000)}. The resulting object-to-object scatter is 
substantial, but there is no systematic bias \citep{Collin(2006)}.
We adopt the prescription used by \citet{Vestergaard(2006)} to compute
black hole masses directly from observables,
\begin{equation}
M_{\mathrm{bh}}=5.4\times10^6\left\{
\left[\frac{\mathrm{\sigma_l(\mathrm{\ion{C}{iv}})}}{1000\mathrm{km/s}}\right]^2
\left[\frac{\lambda L_{\lambda}(1350\,\mathrm{\AA})}{10^{44}\,\mathrm{erg/s}}\right]^{0.53}\right\}
\label{CIV}
\end{equation}
where $L_{\lambda}(1350\,\mathrm{\AA})$ is the monochromatic luminosity at
1350~\AA, and $\sigma_l(\mathrm{\ion{C}{iv}})$ is the dispersion (second central
moment) of the \ion{C}{iv} $\lambda$1550 emission line.

\begin{figure}
\begin{center}
  \includegraphics[clip,angle=-90,width=234pt]{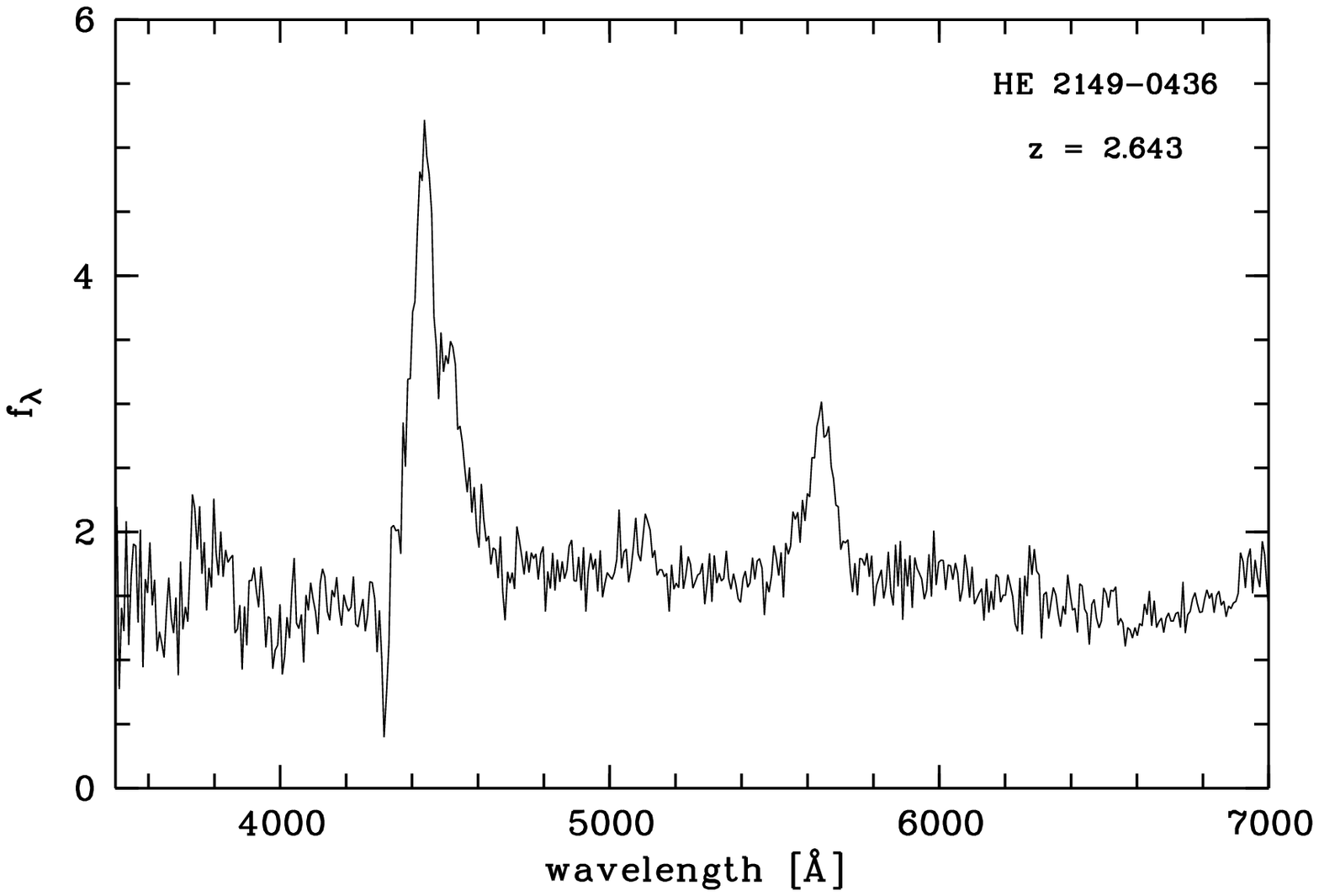}
  \includegraphics[clip,angle=-90,width=234pt]{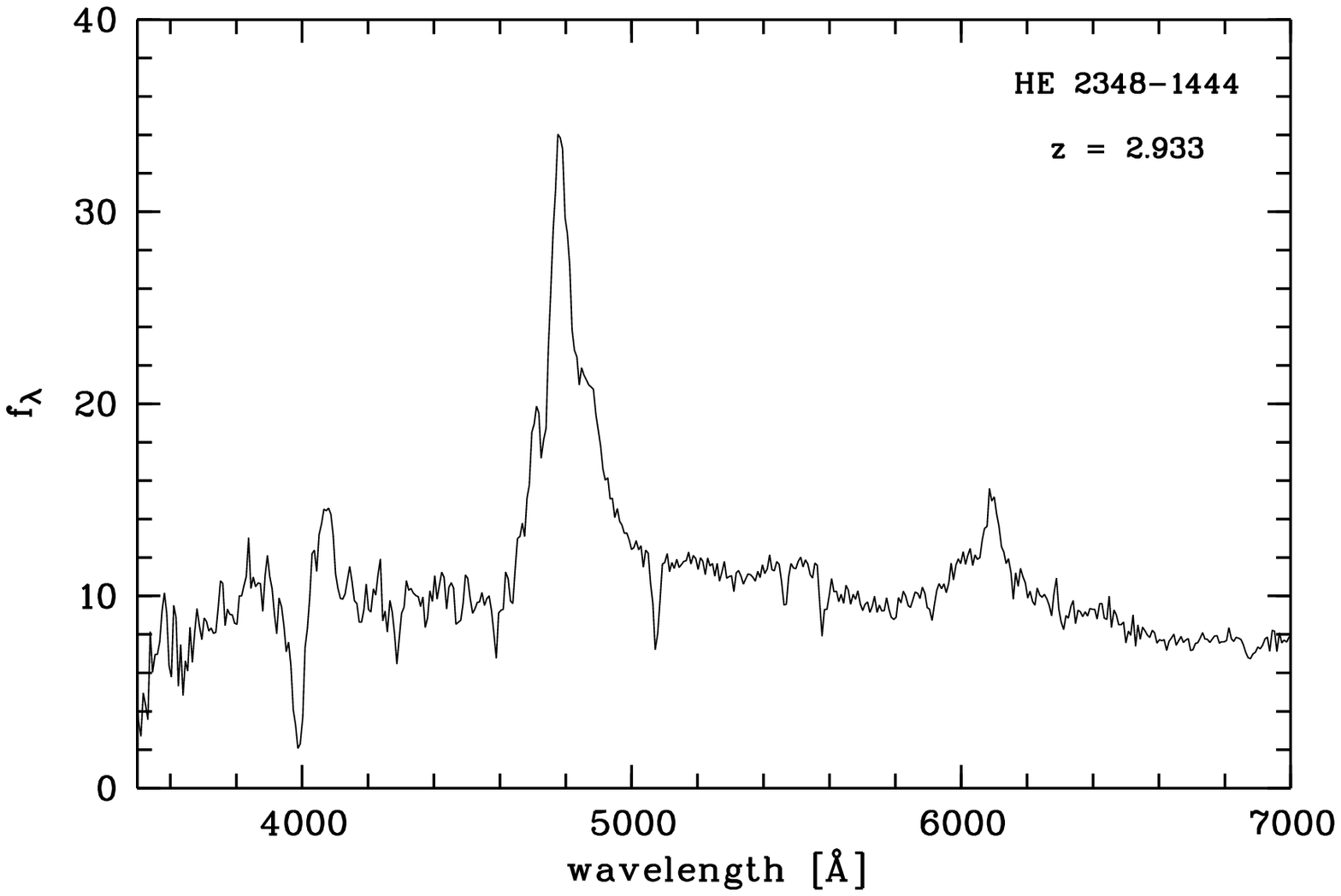}
  \includegraphics[clip,angle=-90,width=234pt]{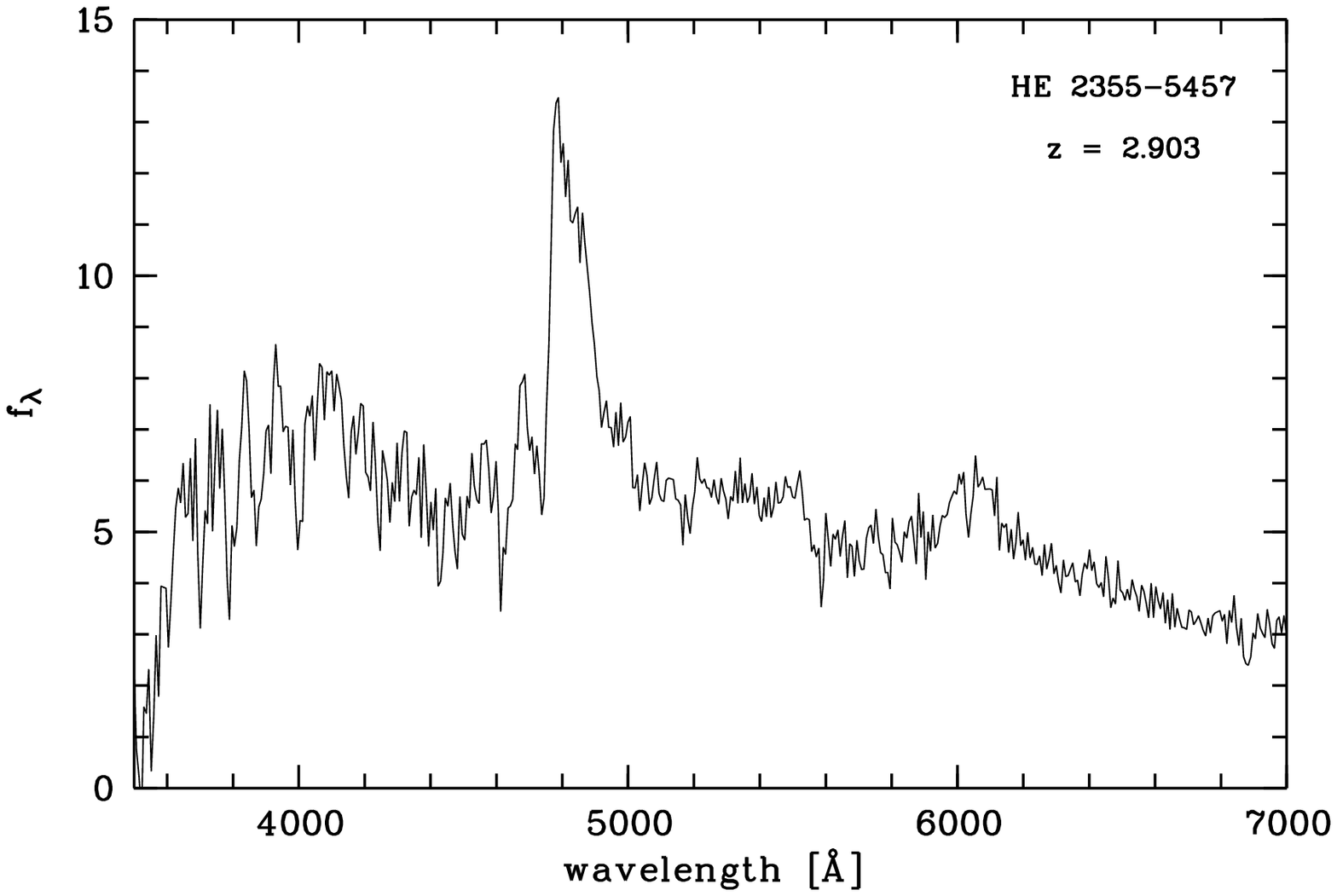}
 \caption{Slit spectra of HE~2149--0436, HE~2355--5457 and HE~2348--1444. 
   $f_\lambda$ is in units of $10^{-16}~\mathrm{erg}\,\mathrm{cm}^{-2}\,\mathrm{s}^{-1}\,\mathrm{\AA}^{-1}$.}
\label{spectra}
\end{center}
\end{figure}

From the follow-up spectroscopy within the Hamburg/ESO Survey we have slit
spectra available of all three QSOs where we resolved the host galaxies.
These spectra are presented in Fig. \ref{spectra}.  We measured the
\ion{C}{iv} line dispersion in the following way: We first subtracted a local
continuum, estimated as a straight line fitted to the line-free windows at
$1440~\mathrm{\AA}<\lambda<1480~\mathrm{\AA}$ and
$1660~\mathrm{\AA}<\lambda<1700~\mathrm{\AA}$; a slight extrapolation of this
provided also the continuum flux at 1350~\AA. We then determined the second
central moment $\mathrm{\sigma_l}$ for each line. Two of our objects (HE~2348--1444 and
HE~2355--5457) show evidence for associated absorption in the blue wing of
\ion{C}{iv} which we had to correct for. In HE~2355--5457 it was
straightforward to interpolate over the absorption line, whereas in
HE~2348--1444 some guesswork was required, given the low spectral resolution.
Fortunately, $\mathrm{\sigma_l}$ is not very sensitive to such manipulations (unlike
the FWHM), and the
difference between the best-guess corrected and an even completely uncorrected
$\mathrm{\sigma_l}$ was only $\sim 10$~\%, which would translate into a 20~\% change in
$M_\mathrm{bh}$. As some degree of correction was undoubtedly needed, we argue
that the additional error introduced this way is smaller.  The results are
presented in Table \ref{T:masses}.

\begin{figure}
\begin{center}
  \includegraphics[clip,angle=-90,width=220pt]{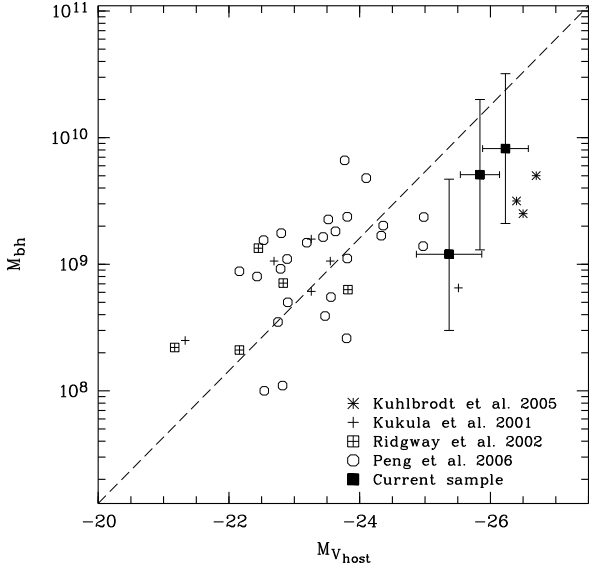}
 \caption{Relation between black hole masses and host luminosities.
   Filled squares are our data; additional data points are taken from the
 literature. The dotted line shows the zero-redshift relation
 from \citet{Dunlop(2003)}.}
\label{mbh_mvh}
\end{center}
\end{figure}

As expected given the high luminosities, the black holes are very massive with
$M_\mathrm{bh}$ up to a few times $10^{9}\:M_{\odot}$. Yet, these masses are 
well within the upper envelopes of the distribution obtained by 
\citet{Vestergaard(2004)} for high-luminosity high-redshift QSOs. 
In Fig.~\ref{mbh_mvh} we investigate where our objects fall within the
$M_{\rm{bh}}-L_V$ relationship, compared to other high redshift QSOs with
detected host galaxies and available black hole mass estimates.
(In this context we use the term $L_V$ to avoid confusion between masses and
absolute magnitudes; nevertheless we always express luminosities in terms
of absolute magnitudes $M_V$.)
Our three objects are perfectly consistent with an approximately linear
$M_{\rm{bh}}-\log L_V$ relation. 

A revised prescription of the $M_{\rm{bh}}-L_V$ relation for low-redshift
bulges and ellipticals was derived by \citet{Dunlop(2003)} as
$M_{\rm{bh}}=1.3\times 10^{-6}L_{\rm{bulge}}^{1.31}$, in solar units.  We
overplot this relation as a dashed line in Fig.~\ref{mbh_mvh}. There is a
surprisingly good agreement between the low-$z$ relation and the high-$z$
data, which at face value might indicate no evolution at all. Of course, this
cannot be true.

It is important to recall that applying the above formula explicitly assumes a
well-defined mass-to-light ratio for QSO host galaxies, essentially that of
old elliptical galaxies at $z\sim 0$ \citep{Jorgensen(1996)}. Overplotting
this relation into a diagram showing data obtained at $z>1.5$ up to $z\sim 3$,
without accounting at least for passive evolution, is therefore misleading.
Only spectral information can reveal the amount of galaxy evolution required
to match the high-redshift results to the local universe.  Unfortunately, for
most of the data shown in Fig.~\ref{mbh_mvh} there are only single-band
measurements available. But since a substantial amount of evolution is
unavoidable, \citet{Peng(2006)} argued that even after conservatively
correcting for only passive evolution with an assumed formation redshift of
$z=5$, there must be a significant offset between the local and high-redshift
relations. They concluded that at redshifts $z\sim 2$, the ratio of black hole
to stellar bulge masses must have been higher than `today' by at least a
factor of 4. We explore in the next subsection how our colour measurements fit
into this picture.

Using our black hole masses we can also predict Eddington luminosities
and Eddington ratios $\epsilon$ (cf.\ Table~\ref{T:masses}). We find
that even these high-luminosity QSOs do not exceed 
Eddington-limited accretion, in fact their inferred Eddington ratios 
are completely consistent with those obtained for less luminous AGN
at all redshifts \citep[e.g.,][]{Kollmeier(2006)}.

\begin{table*}
\begin{center}
\begin{tabular}{lrrrrrrrr}
\hline\noalign{\smallskip}
Object & $H-\ks$ (obs.)  &  $B-V$ (rest) & $M_\star$ & $\mathrm{\sigma_l}$     &  log$\left(\lambda L_\lambda\right)$   & $M_{\mathrm{bh}}$ &  $\epsilon$ &  $\mu$ \\
       &                 &               & [$10^{11}M_{\odot}$] &      km/s                & erg/s   &[$10^{9}M_{\odot}$]&            & \\
\noalign{\smallskip}\hline\noalign{\smallskip}
HE~2149$-$0436 &  $>$ 0.03& $>-$0.33 & $>$1.1               & 2900$\pm150$ & 46.7 &  1.2 &  0.3 &     $<$0.011 \\
HE~2355$-$5457 &  0.26    & $-$0.16  & ~2.6$^{+5.5}_{-1.8}$ & 4100$\pm400$ & 47.3 &  5.1 &  0.3 &      0.019$^{+0.064}_{-0.012}$ \\                        
HE~2348$-$1444 &  0.39    & $-$0.07  & ~4.9$^{+10.0}_{-3.3}$& 4300$\pm300$ & 47.6 &  8.2 &  0.6 &      0.017$^{+0.055}_{-0.011}$ \\
\noalign{\smallskip}\hline
\end{tabular}
\end{center}
\caption[]{Host galaxy and nuclear properties: Observed $H-\ks$,
    rest-frame $B-V$, stellar mass $M_\star$, $\mathrm{\ion{C}{iv}}$ emission
    line dispersion $\mathrm{\sigma_l}$,$\mathrm{\ion{C}{iv}}$
    continuum fluxes at 1350\AA, estimated black hole mass
    $M_{\mathrm{bh}}$, Eddington ratio $\epsilon$ and the black hole / bulge mass ratio $\mu$. 

}

\label{T:masses}
\end{table*}

\subsection{Host galaxy colours: Evidence for young stellar populations}

For the two QSO hosts detected in both $H$ and \ks\ we could also derive
rest-frame $B-V$ colours; for the third object we obtained a conservative
limit on $B-V$. We find that the colours of our two objects depend
very little on the assumed scale lengths or morphology. The results are
compiled in Table~\ref{T:masses}.  The host galaxies are remarkably blue, with
a $B-V$ of just below 0 (in the Vega system) with an uncertainty of
0.3 mag, including our individual uncertainties in the host magnitudes of about
0.2 mag and in the K-correction of about 0.05 mag) - thus
corresponding to the colours of A-type stars. In terms of luminosity-weighted
stellar ages (the ages of single-burst stellar populations (SSP) with the same
$B-V$), we obtain values of $\sim 300$~Myrs for both objects.

Finding that QSO host galaxies at $z\sim 3$ have blue colours and young
stellar populations is not really surprising, but it nevertheless poses
relevant constraints. Note that a literal interpretation of the SSP ages 
leads to a formation redshift of $z\simeq 3.3$, considerably lower
than the $z=5$ formation (conservatively) assumed by \cite{Peng(2006)}. 

Blue colours of QSO host galaxies have recently been established as ubiquitous
at lower redshifts. In \citet{Jahnke(2004)} we demonstrated that $z<0.2$ quasar
hosts with elliptical morphology have significant bluer colours than their
inactive counterparts. \citet{Sanchez(2004)} presented a similar conclusion
for a sample of QSO hosts at $z\simeq 0.7$ using HST imaging. In a different
approach, \citet{Kauffmann(2003)} used SDSS data to show that Seyfert-2
galaxies are typically hosted by massive early-type galaxies with strong
evidence for recent star formation. Comparing our new rest-frame $B-V$ values 
with those typically found at low redshifts, we find them to be even somewhat 
bluer but formally still consistent (given our error bars).

Corresponding data on high redshift QSO hosts are very scarce. From optical
imaging of several $z>1.8$ QSOs with HST, \cite{Jahnke(2004)2} found evidence
for substantial young stellar populations, with rest-frame UV luminosities
comparable to Lyman Break Galaxies. We do not sample the rest-frame UV in
our objects (and presumably would be overwhelmed by the nuclei anyway), so a
direct comparison is not possible. We conclude that our finding
of blue colours is qualitatively in very good agreement with most previous 
results.

\begin{figure}
\begin{center}
  \includegraphics[clip,angle=-90,width=224pt]{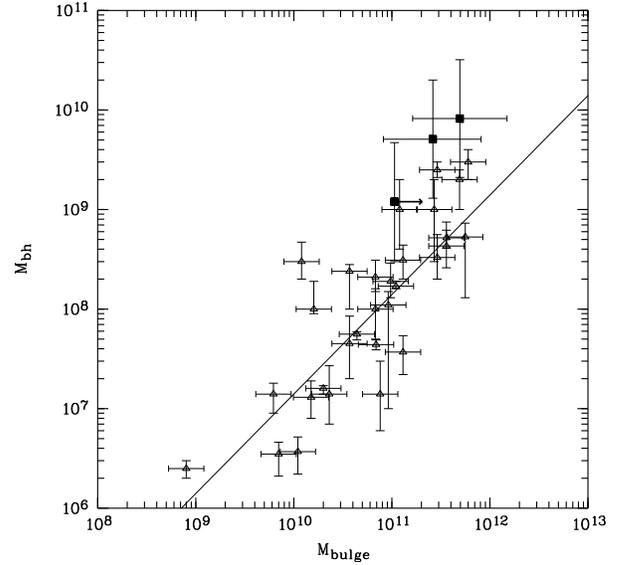}
  \caption{$M_\mathrm{bh}$-$M_\mathrm{bulge}$ relation reproduced from
    \citet{Haering(2004)} for a sample of local galaxies. The solid line
    represents a constant mass ratio $M_{\mathrm{bh}}/M_{\mathrm{bulge}} =
    0.0014$. The three filled symbols show our 
    estimated stellar and virial black hole masses for the
    three $z\sim 3$ QSOs.
    }
\label{mbh_mbulge}
\end{center}
\end{figure}

\subsection{Stellar mass-to-light ratios and the $M_\mathrm{bh}$-$M_\mathrm{bulge}$ relation}

We can now return to improve our handling of the relation between black holes
and their host galaxies. While there is no unique conversion of a single
colour measurement into a stellar mass-to-light ratio, it is clear that the
blue $B-V$ colours found for our objects imply rather low $M/L$.
\citet{Bell(2003)} demonstrated that for a variety of star formation
histories, stellar $M/L$ can be quite robustly predicted from $B-V$, and we
used their formula
\begin{equation}
\mathrm{log}_{10}(M/L_V)=-0.628+(1.305\times (B-V))\:.
\label{Eq_stell_mass}
\end{equation}
where $M/L_V$ is given in solar units.

Using the $B-V$ colours of Table~\ref{T:masses}, we obtained $M/L_V$
  ratios in the range of 0.05--0.2. Given our estimated uncertainties for the
  colours the associated uncertainties in $M/L_V$ are $\pm0.25$~dex  
 
Combining our $M/L$ ratio estimates with the host galaxy luminsities above leads
to stellar masses in the range of a few $10^{11}\:M_{\odot}$ for the two main
objects. The formally computed values are listed in Table~\ref{T:masses}.
Thus, these galaxies are still very massive, but well within the range of the
empirically samples stellar mass function of inactive galaxies as determined
recently by \citet{Fontana(2006)}.

We then combined the black hole masses derived above with our estimated
stellar masses. In doing so we assumed our luminous, massive host galaxies to
be dominated by a bulge-like component. Our derived ratio $\mu$ between
$M_\mathrm{bh}$ and $M_\mathrm{bulge}$ is given in Table~\ref{T:masses}; 
the two values and one upper limit are all $> 0.01$. Formally, this is
considerably larger than the local ratio of $\sim 0.0015$ obtained for
inactive galaxies, although the discrepancy is only of the order of
1--2$\sigma$ given the substantial individual error bars.  Note that any
additional non-bulge component would increase the discrepancy.

We illustrate our results in Fig.~\ref{mbh_mbulge} by overplotting our high-redshift
data points into the local $M_\mathrm{bh}$-$M_\mathrm{bulge}$ relation obtained by
\citet{Haering(2004)}, together with a constant $\mu = 0.0014$ relation (solid
line). Our points are always above the local relation, although only marginally.
Furthermore, the masses probed by our high-luminosity objects are at the
extreme end of the range probed by the local data, and it is not yet clear that 
it can be extrapolated at constant $\mu$ into the high mass regime.

Measuring $M_\mathrm{bh}/M_\mathrm{bulge}$ for different cosmological epochs,
is presently a hot topic, and many groups have now endeavoured to estimate
$\mu$ with a variety of different methods. Interestingly, most find evidence
for $\mu$ to increase with lookback time. We already mentioned
\citet{Peng(2006),Peng(2006)3} who obtained $\mu_{z=2}/\mu_{z=0} \ga 4$ at
$z\simeq 2$, with the inequality depending on the unknown stellar
mass-to-light ratios of their galaxies. A similar result was obtained by
\citet{McLure(2006)}, combining $M_\mathrm{bulge}$ of radio galaxies with
$M_\mathrm{bh}$ of radio-loud QSOs, assuming these to be actually the same
class of objects.  An extremely large evolution in $\mu(z)$ was claimed by
\citet{Shields(2006)1} who used CO emission line widths as surrogates for
stellar velocity dispersion and found that $\mu_{z=4}/\mu_{z=0}\sim 50$ at
$z\simeq 4$. On the other hand, \citet{Lauer(2007)} recently
demonstrated that comparing high-redshift quasar measurements with
low-redshift galaxies can lead to systematically overestimating the amount of
evolution in $\mu$. Our experimental setup is clearly affected by such biases,
as the quasars were drawn from a survey for very luminous quasars. From Fig.~6
in \citet{Lauer(2007)} we estimate that our value of $\mu$ may be biased by up
to 0.5~dex towards high values. Taking this bias into account then there is
still a systematic offset of our $\mu$ values to the local one, but it is now
well below our $1\sigma$ errors.

\section{Conclusions}

We have presented rest-frame $B$ and $V$ imaging of QSO host galaxies,
allowing us for the first time to measure an optical $B-V$ colour for at least
some of our targets. We found these galaxies to show rather, but not extremely
blue colours, indicative of either an enormous burst of star formation a few
hundred Myr ago, or of ongoing star formation on a much lower but still
substantial level. Clearly, a single colour cannot differentiate between these
options. It might be useful to search for CO emission in these QSOs which,
if found, could be taken as evidence for ongoing star formation. 

At least one of our QSOs is embedded in a highly asymmetric host galaxy
structure, indicative of gravitational interaction or merging. It is
interesting to note that this asymmetric structure is more prominent in the
$H$ (rest-frame $B$) than in the \ks\ (rest-frame $V$) band image, which
suggests that the blue light from young stars comes preferentially from this
structure.

Our observations show that high-luminosity QSOs ($M_B \la -28$) at $z\sim 3$
already live in very massive galaxies of several times $10^{11}\:M_\odot$.
Such massive structures are rare at high $z$, but not exceedingly rare:
Following \citet{Fontana(2006)}, the space density of galaxies with
stellar masses above $3\times 10^{11}\:M_\odot$ at $z\sim 3$ is of the order 
of $10^{-5}$--$10^{-6}$~Mpc$^{-3}$, still substantially higher than the 
number density of luminous QSOs \citep{Wisotzki(2000)2}. Thus, even at the
peak of the cosmic QSO activity, only a small fraction of galaxies is 
growing their black holes at near Eddington rates.

By combining stellar bulge masses and virial black hole masses for two
objects, we could estimate the ratio $\mu$ of these quantities in individual
objects. We found a marginal excess of $\mu(z=3)$ compared to the local
relation. Taking into account the substantial uncertainties, and considering
recent analyses of systematic biases, however, our results are still
consistent with no evolution in the black hole mass - bulge mass ratio.
Still we believe that the approach as such is promising, and with larger
samples and more accurate measurements, obtained at several redshifts,
it has the potential to significantly constrain the degree of co-evolution
between black holes and hosting bulges.

\begin{acknowledgements}
We thank Dr.\ Asmus Boehm, Dr. Isabelle Gavignaud and Dr.\ Chien Y. Peng for
illuminating discussions. We are gratful to the anonymous referee who made 
numerous suggestions how to improve this paper.
\end{acknowledgements}

\bibliographystyle{aa}
\bibliography{reference}

\begin{thebibliography}{37}
\expandafter\ifx\csname natexlab\endcsname\relax\def\natexlab#1{#1}\fi

\bibitem[{{Abraham} {et~al.}(1992){Abraham}, {Crawford}, \&
  {McHardy}}]{Abraham(1992)}
{Abraham}, R.~G., {Crawford}, C.~S., \& {McHardy}, I.~M. 1992, \apj, 401, 474

\bibitem[{{Bell} {et~al.}(2003){Bell}, {McIntosh}, {Katz}, \&
  {Weinberg}}]{Bell(2003)}
{Bell}, E.~F., {McIntosh}, D.~H., {Katz}, N., \& {Weinberg}, M.~D. 2003, ApJS,
  149, 289

\bibitem[{{Collin} {et~al.}(2006){Collin}, {Kawaguchi}, {Peterson}, \&
  {Vestergaard}}]{Collin(2006)}
{Collin}, S., {Kawaguchi}, T., {Peterson}, B.~M., \& {Vestergaard}, M. 2006,
  \aap, 456, 75

\bibitem[{{Dunlop} {et~al.}(2003){Dunlop}, {McLure}, {Kukula}, {Baum}, {O'Dea},
  \& {Hughes}}]{Dunlop(2003)}
{Dunlop}, J.~S., {McLure}, R.~J., {Kukula}, M.~J., {et~al.} 2003, MNRAS, 340,
  1095

\bibitem[{{Falomo} {et~al.}(2004){Falomo}, {Kotilainen}, {Pagani}, {Scarpa}, \&
  {Treves}}]{Falomo(2004)}
{Falomo}, R., {Kotilainen}, J.~K., {Pagani}, C., {Scarpa}, R., \& {Treves}, A.
  2004, \apj, 604, 495

\bibitem[{{Falomo} {et~al.}(2005){Falomo}, {Kotilainen}, {Scarpa}, \&
  {Treves}}]{Falomo(2005)}
{Falomo}, R., {Kotilainen}, J.~K., {Scarpa}, R., \& {Treves}, A. 2005, \aap,
  434, 469

\bibitem[{{Floyd} {et~al.}(2004){Floyd}, {Kukula}, {Dunlop}, {McLure},
  {Miller}, {Percival}, {Baum}, \& {O'Dea}}]{Floyd(2004)}
{Floyd}, D.~J.~E., {Kukula}, M.~J., {Dunlop}, J.~S., {et~al.} 2004, MNRAS, 355,
  196

\bibitem[{{Fontana} {et~al.}(2006){Fontana}, {Salimbeni}, {Grazian},
  {Giallongo}, {Pentericci}, {Nonino}, {Fontanot}, {Menci}, {Monaco},
  {Cristiani}, {Vanzella}, {de Santis}, \& {Gallozzi}}]{Fontana(2006)}
{Fontana}, A., {Salimbeni}, S., {Grazian}, A., {et~al.} 2006, \aap, 459, 745

\bibitem[{{H{\"a}ring} \& {Rix}(2004)}]{Haering(2004)}
{H{\"a}ring}, N. \& {Rix}, H.-W. 2004, \apjl, 604, L89

\bibitem[{{Jahnke} {et~al.}(2004{\natexlab{a}}){Jahnke}, {Kuhlbrodt}, \&
  {Wisotzki}}]{Jahnke(2004)}
{Jahnke}, K., {Kuhlbrodt}, B., \& {Wisotzki}, L. 2004{\natexlab{a}}, MNRAS,
  352, 399

\bibitem[{{Jahnke} {et~al.}(2004{\natexlab{b}}){Jahnke}, {S{\'a}nchez},
  {Wisotzki}, {Barden}, {Beckwith}, {Bell}, {Borch}, {Caldwell},
  {H{\"a}ussler}, {Heymans}, {Jogee}, {McIntosh}, {Meisenheimer}, {Peng},
  {Rix}, {Somerville}, \& {Wolf}}]{Jahnke(2004)2}
{Jahnke}, K., {S{\'a}nchez}, S.~F., {Wisotzki}, L., {et~al.}
  2004{\natexlab{b}}, ApJ, 614, 568

\bibitem[{{Jorgensen} {et~al.}(1996){Jorgensen}, {Franx}, \&
  {Kjaergaard}}]{Jorgensen(1996)}
{Jorgensen}, I., {Franx}, M., \& {Kjaergaard}, P. 1996, MNRAS, 280, 167

\bibitem[{{Kaspi} {et~al.}(2000){Kaspi}, {Smith}, {Netzer}, {Maoz}, {Jannuzi},
  \& {Giveon}}]{Kaspi(2000)}
{Kaspi}, S., {Smith}, P.~S., {Netzer}, H., {et~al.} 2000, \apj, 533, 631

\bibitem[{{Kauffmann} {et~al.}(2003){Kauffmann}, {Heckman}, {Tremonti},
  {Brinchmann}, {Charlot}, {White}, {Ridgway}, {Brinkmann}, {Fukugita}, {Hall},
  {Ivezi{\'c}}, {Richards}, \& {Schneider}}]{Kauffmann(2003)}
{Kauffmann}, G., {Heckman}, T.~M., {Tremonti}, C., {et~al.} 2003, MNRAS, 346,
  1055

\bibitem[{{Kollmeier} {et~al.}(2006){Kollmeier}, {Onken}, {Kochanek}, {Gould},
  {Weinberg}, {Dietrich}, {Cool}, {Dey}, {Eisenstein}, {Jannuzi}, {Le Floc'h},
  \& {Stern}}]{Kollmeier(2006)}
{Kollmeier}, J.~A., {Onken}, C.~A., {Kochanek}, C.~S., {et~al.} 2006, \apj,
  648, 128

\bibitem[{{Kuhlbrodt} {et~al.}(2005){Kuhlbrodt}, {{\"O}rndahl}, {Wisotzki}, \&
  {Jahnke}}]{Kuhlbrodt(2005)}
{Kuhlbrodt}, B., {{\"O}rndahl}, E., {Wisotzki}, L., \& {Jahnke}, K. 2005, AAP,
  439, 497

\bibitem[{{Kuhlbrodt} {et~al.}(2004){Kuhlbrodt}, {Wisotzki}, \&
  {Jahnke}}]{Kuhlbrodt(2004)}
{Kuhlbrodt}, B., {Wisotzki}, L., \& {Jahnke}, K. 2004, MNRAS, 349, 1027

\bibitem[{{Kukula} {et~al.}(2001){Kukula}, {Dunlop}, {McLure}, {Miller},
  {Percival}, {Baum}, \& {O'Dea}}]{Kukula(2001)}
{Kukula}, M.~J., {Dunlop}, J.~S., {McLure}, R.~J., {et~al.} 2001, MNRAS, 326,
  1533

\bibitem[{{Lauer} {et~al.}(2007){Lauer}, {Tremaine}, {Richstone}, \&
  {Faber}}]{Lauer(2007)}
{Lauer}, T.~R., {Tremaine}, S., {Richstone}, D., \& {Faber}, S.~M. 2007,
  arXiv:astro-ph/0705.4103

\bibitem[{{Magorrian} {et~al.}(1998){Magorrian}, {Tremaine}, {Richstone},
  {Bender}, {Bower}, {Dressler}, {Faber}, {Gebhardt}, {Green}, {Grillmair},
  {Kormendy}, \& {Lauer}}]{Magorrian(1998)}
{Magorrian}, J., {Tremaine}, S., {Richstone}, D., {et~al.} 1998, AJ, 115, 2285

\bibitem[{{McLeod} \& {Rieke}(1995)}]{McLeod(1995)}
{McLeod}, K.~K. \& {Rieke}, G.~H. 1995, ApJ, 441, 96

\bibitem[{{McLeod} {et~al.}(1999){McLeod}, {Rieke}, \&
  {Storrie-Lombardi}}]{McLeod(1999)}
{McLeod}, K.~K., {Rieke}, G.~H., \& {Storrie-Lombardi}, L.~J. 1999, \apjl, 511,
  L67

\bibitem[{{McLure} \& {Jarvis}(2002)}]{McLure(2002)}
{McLure}, R.~J. \& {Jarvis}, M.~J. 2002, \mnras, 337, 109

\bibitem[{{McLure} {et~al.}(2006){McLure}, {Jarvis}, {Targett}, {Dunlop}, \&
  {Best}}]{McLure(2006)}
{McLure}, R.~J., {Jarvis}, M.~J., {Targett}, T.~A., {Dunlop}, J.~S., \& {Best},
  P.~N. 2006, \mnras, 368, 1395

\bibitem[{{McLure} {et~al.}(1999){McLure}, {Kukula}, {Dunlop}, {Baum}, {O'Dea},
  \& {Hughes}}]{McLure(1999)}
{McLure}, R.~J., {Kukula}, M.~J., {Dunlop}, J.~S., {et~al.} 1999, MNRAS, 308,
  377

\bibitem[{{Peng} {et~al.}(2006{\natexlab{a}}){Peng}, {Impey}, {Ho}, {Barton},
  \& {Rix}}]{Peng(2006)}
{Peng}, C.~Y., {Impey}, C.~D., {Ho}, L.~C., {Barton}, E.~J., \& {Rix}, H.-W.
  2006{\natexlab{a}}, \apj, 640, 114

\bibitem[{{Peng} {et~al.}(2006{\natexlab{b}}){Peng}, {Impey}, {Rix},
  {Kochanek}, {Keeton}, {Falco}, {Leh{\'a}r}, \& {McLeod}}]{Peng(2006)3}
{Peng}, C.~Y., {Impey}, C.~D., {Rix}, H.-W., {et~al.} 2006{\natexlab{b}}, \apj,
  649, 616

\bibitem[{{Persson} {et~al.}(1998){Persson}, {Murphy}, {Krzeminski}, {Roth}, \&
  {Rieke}}]{Persson(1998)}
{Persson}, S.~E., {Murphy}, D.~C., {Krzeminski}, W., {Roth}, M., \& {Rieke},
  M.~J. 1998, \aj, 116, 2475

\bibitem[{{Peterson} \& {Wandel}(2000)}]{Peterson(2000)}
{Peterson}, B.~M. \& {Wandel}, A. 2000, \apjl, 540, L13

\bibitem[{{Ridgway} {et~al.}(2001){Ridgway}, {Heckman}, {Calzetti}, \&
  {Lehnert}}]{Ridgway(2001)}
{Ridgway}, S.~E., {Heckman}, T.~M., {Calzetti}, D., \& {Lehnert}, M. 2001, ApJ,
  550, 122

\bibitem[{{S{\'a}nchez} {et~al.}(2004){S{\'a}nchez}, {Jahnke}, {Wisotzki},
  {McIntosh}, {Bell}, {Barden}, {Beckwith}, {Borch}, {Caldwell},
  {H{\"a}ussler}, {Jogee}, {Meisenheimer}, {Peng}, {Rix}, {Somerville}, \&
  {Wolf}}]{Sanchez(2004)}
{S{\'a}nchez}, S.~F., {Jahnke}, K., {Wisotzki}, L., {et~al.} 2004, ApJ, 614,
  586

\bibitem[{{Shields} {et~al.}(2006){Shields}, {Menezes}, {Massart}, \& {Vanden
  Bout}}]{Shields(2006)1}
{Shields}, G.~A., {Menezes}, K.~L., {Massart}, C.~A., \& {Vanden Bout}, P.
  2006, \apj, 641, 683

\bibitem[{{Vestergaard}(2002)}]{Vestergaard(2002)}
{Vestergaard}, M. 2002, \apj, 571, 733

\bibitem[{{Vestergaard}(2004)}]{Vestergaard(2004)}
{Vestergaard}, M. 2004, in ASP Conf. Ser. 311: AGN Physics with the Sloan
  Digital Sky Survey, 69--+

\bibitem[{{Vestergaard} \& {Peterson}(2006)}]{Vestergaard(2006)}
{Vestergaard}, M. \& {Peterson}, B.~M. 2006, \apj, 641, 689

\bibitem[{{Wisotzki}(2000)}]{Wisotzki(2000)2}
{Wisotzki}, L. 2000, \aap, 353, 853

\bibitem[{{Wisotzki} {et~al.}(2000){Wisotzki}, {Christlieb}, {Bade},
  {Beckmann}, {K{\"o}hler}, {Vanelle}, \& {Reimers}}]{Wisotzki(2000)}
{Wisotzki}, L., {Christlieb}, N., {Bade}, N., {et~al.} 2000, AAP, 358, 77

\end{thebibliography}

\end{document}